\documentclass[12pt,a4paper]{article}
\pdfoutput=1
\usepackage{jheppub}
\usepackage[utf8]{inputenc}
\usepackage[T1]{fontenc}

\usepackage{amsmath,physics,float}  
\usepackage{amssymb}  
\usepackage{latexsym}
\usepackage{graphicx,xcolor}

\usepackage{amsfonts}
\usepackage{braket}
\usepackage{mathrsfs}

\usepackage{subfig} 
\usepackage{empheq}

\usepackage[shortlabels]{enumitem}

\usepackage{pgfplots}
\usepackage{natbib}

\usepackage{bigints}

\newcommand{\beq}{\begin{equation}}
\newcommand{\eeq}{\end{equation}}

\newcommand{\lc}{\left(}
\newcommand{\rc}{\right)}

\begin{document}

\title{\boldmath Small corrections to semiclassical gravity and their role in information paradoxes}

\author{Joydeep Chakravarty}
\emailAdd{joydeep.chakravarty@icts.res.in}

\affiliation[]{International Centre for Theoretical Sciences (ICTS-TIFR)\\
Tata Institute of Fundamental Research\\
Shivakote, Hesaraghatta,
Bangalore 560089, India.}
\vspace{4pt}

\begin{abstract}
{In this essay, we qualitatively demonstrate how small non-perturbative corrections are a necessary addition to semiclassical gravity's path integral. We use this to discuss implications for Hawking's information paradox and the bags of gold paradox.\\
\leftline{}
\centerline{(\textit{Essay written for the Gravity Research Foundation 2021 Awards for Essays on Gravitation.})}}
\end{abstract} 

\maketitle

\section{Introduction}
This work's primary goal is to shed light on the origin of small non-perturbative corrections to semiclassical gravity's path integral, and how they can be helpful in understanding certain information paradoxes involving the black hole interior \cite{Maldacena:2001kr, Barbon:2004ce, Mathur:2012np}. The focus of this work is to address these issues from the bulk viewpoint, complementary to interesting developments summarized in \cite{Mathur:2012np, Almheiri:2020cfm, Raju:2020smc} and references therein. To understand the necessity of small corrections, we restrict ourselves to asymptotically AdS spacetime, where such developments are pretty concretely understood \cite{Maldacena:1997re, Witten-ads-and-holography, Gubser:1998bc}. 

Let us briefly discuss what exactly we are interested in, beginning with Hawking's information paradox. By studying quantum fields over a fixed black hole background, it can be shown that the outgoing Hawking radiation is thermal, a feature that arises due to the presence of the horizon. This leads to the following question: how is the time evolution of an initially pure state, say a collapsing star, to the final thermal state of Hawking radiation consistent with quantum mechanics (QM)? The second situation we are interested in regards bags of gold spacetimes \cite{Marolf:2008tx}. Within the framework of effective field theory (EFT), we can create a vast number of spatially separated semiclassical excitations in the black hole interior and thus arrive at an entropy much larger than the Bekenstein Hawking entropy. Besides, such configurations also violate black holes' expected spectral properties \cite{Chakravarty:2020wdm, CR}.

It has long been argued that Hawking's version of the paradox is not well-posed, as small corrections to a thermal density matrix can purify it, a fact which was overlooked in Hawking's original calculation \cite{Raju:2020smc}. A more robust reformulation of this statement purely within a statistical framework is that correlators over typical states in a statistical ensemble mimic thermal correlators upto small corrections \cite{lloyd2013pure}. In this work, we demonstrate the origin of such small corrections and how they provide a basis for these statistical statements. We also revisit the arguments in \cite{Chakravarty:2020wdm, CR} why the same small corrections play a role in resolving the bags of gold paradox in a different light.

\section{Semiclassical gravity path integral}
To explicitly demonstrate small corrections, it is convenient to rewrite the semiclassical path integral as a path integral over coherent states. To build-up, let us first look at coherent state path integrals in QM.

\subsection{Coherent state path integrals in QM}
Consider a one-dimensional simple harmonic oscillator. Let $\ket{z}$ be the coherent state defined by $\ket{z} = e^{za^{\dagger}}\ket{0}$, where $\ket{0}$ denotes the vacuum. The inner product between two coherent states is given by $\braket{z|z'} = e^{\bar{z} z'}$. The lifting of functions from the phase space to operators acting on the Hilbert space is performed using the Sudarshan-Mehta P representation:
\begin{equation} \label{f}
 \hat{f} = \int \frac{d^2z}{2\pi i}\, e^{-z\bar{z}} f(z) \ket{z}\bra{z}   
\end{equation}
Defining $\braket{z|\psi} = \psi(\bar{z})$, we derive the coherent state path integral by inserting the identity operator given by \eqref{f} in the discretization of the matrix element $\braket{f|e^{-iTH}|i}$, where $H$ denotes the Hamiltonian. The formal expression for the path integral is given by
\begin{equation} \label{ffs}
    \braket{f|\,e^{-iH(t_f- t_i)}\,|i} = \int \mathcal{D}z \mathcal{D}\bar{z}\, \,  e^{i S + \frac{1}{2}\left( \abs{z_i}^2 + \abs{z_f}^2\right)}\, \bar{\psi}_f(z_f) \, \psi_i(\bar{z}_i)
\end{equation}
 where $S$ denotes the action $S = \int_{t_i}^{t_f} dt \lc \frac{1}{2i} \lc z \, \partial_t \bar{z} - \bar{z}\, \partial_t z \rc - H(z,\bar{z}) \rc$. The other terms in \eqref{ffs} arise by expressing the initial and the final states in the basis of coherent states. We will move onto coherent states in free field theory and finally to the case with gravity turned on in the next subsection.
\subsection{Coherent state path integrals in free field theory and gravity}
Within EFT, we can similarly define normalized coherent states over a vacuum $\ket{\Omega}$ by 
$$\ket{\vec{z}} = \exp\Big( -\sum_i \frac{\abs{z_i}^2}{2} +  a_i^{\dagger} z_i\Big) \ket{\Omega},$$
where $a_i$ are annihilation operators. The absolute value of the inner product between two such states is given by $\abs{\braket{z|z'}} = e^{- \frac{\abs{z-z'}^2}{2}}$. We again lift functions from a $2n$-dimensional phase space to operators acting on the Hilbert space using the operator
\begin{equation} \label{f1}
 \hat{f} = \int \frac{d^{2n}z}{(2\pi i)^n} f(z) \ket{z}\bra{z}   ,
\end{equation}
Taking the expectation value of the operator in \eqref{f1} over a coherent state $\ket{\alpha}$, we obtain
\beq \label{f2}
\braket{\alpha\,|\hat{f}\,|\alpha} =  \int \frac{d^{2n}z}{(2\pi i)^n} \, f(z) \, \exp\lc-\abs{z-\alpha}^2 \rc.
\eeq
It is straightforward to use the identity operator from \eqref{f1} to write the coherent state path integral for an EFT Hamiltonian $H$, as done in \eqref{ffs} for the QM system. Analogously, we can also write the coherent states and the path integral for semiclassical gravity as well.

We will work within the linearized gravity approximation, where we fix a background metric $g^0_{\mu \nu}$ and consider small linearized perturbations $h_{\mu\nu}$ about the same. We perform ADM decomposition of the metric and quantize the system by lifting the Poisson algebra between the metric on the spacelike slice and its' conjugate momenta to commutators. Due to linearization, this lifting defines canonical commutators between the creation and the annihilation operators. 

Within this context, we study the coherent states of linearized gravity (again denoted by $\ket{z}$) over a vacuum $\ket{\Omega}$ \cite{Papadodimas:2015jra, Papadodimas:2015xma}. Then the classical values of the metric $g$ and the scalar field $\phi$ are given by $\braket{z|\hat{g}|z} = g$ and $\braket{z|\hat{\phi}|z} = \phi$, where $g = g^0 + h$. We again play the same game for gravity and rewrite the path integral. Let us formally denote the coherent state partition function, which has periodic boundary conditions in imaginary time as
\beq \label{costate}
Z(\beta) = \int \mathcal{D}z \mathcal{D}\bar{z}\, e^{- S(z,\bar{z}, \beta)}. 
\eeq
Here $S(z,\bar{z}, \beta)$ includes the Einstein Hilbert action coupled to the scalar field and contributions arising due to the coherent state construction. 
\section{Origin of small corrections and implications for information paradoxes}
So far, we have performed non-controversial operations within EFT. We now discuss why such a lifting of functions from the phase space to Hilbert space is not correct in gravity. Within ordinary QFT the expected fall-off of coherent states is given by $\abs{\braket{z|z'}} = e^{- \frac{\abs{z-z'}^2}{2}}$ for large enough $\abs{z-z'}$. Thus coherent states with classical expectation values separated far apart in the phase space have vanishingly small overlaps. However, this does not hold for gravity's case. Within the formalism of AdS-CFT, using the dual CFT, it can be shown that states separated far apart in the phase space do not have vanishingly small overlaps as expected, but the overlaps saturate at O$\lc e^{-\mathcal{N}} \rc$, where $G_N \propto \frac{1}{\mathcal{N}}$ \cite{Papadodimas:2015jra, Papadodimas:2015xma, Cotler:2016fpe, Jafferis:2017tiu}. 

We will briefly describe how the gravitational path integral receives small corrections. While discretizing the partition function to rewrite it as a coherent state partition function in \eqref{costate}, we have utilized the identity operator in \eqref{f1}, with $\hat{f} = 1$. However as we just discussed, for $z \gg \alpha$, the inner product between the coherent states saturate at $\abs{\braket{z|\alpha}} =\text{O} \lc e^{-\mathcal{N}} \rc$. Consequently, equation \eqref{f2} needs to be appropriately modified, and thus \eqref{f1} as well. We will now study the implications of these small corrections for black holes.
\subsection{Implications for Hawking's paradox}
Consider the $k$-point correlator given by 
$\braket{\psi_{BH}|\, O(x_1) \,O(x_2) \dots \, O(x_k)|\psi_{BH}},$ where $\ket{\psi_{BH}}$ denotes the state of a black hole. Here we assume that $k \ll \mathcal{N}$ so that the gravitational perturbation theory is well defined, and the operator insertions have energies peaked about $\omega_i$ such that $\omega_i \ll \mathcal{N}$ so as to ignore backreaction. A standard result in EFT follows that $k$-point correlators are thermal:
\beq \label{kpt}
\braket{\psi_{BH}|\, O(x_1) \,O(x_2) \dots \, O(x_k)|\psi_{BH}} =  \text{Tr} \lc \rho \, O(x_1) \,O(x_2) \dots \, O(x_k)\rc
\eeq
where $\rho$ denotes the thermal density matrix. Since in a theory of gravity we do not have local diffeomorphism-invariant operators, we restrict our discussion to \textit{approximately local} operators, defined such that $O(x)$ is smeared over a small region around $x$. When operator insertions at $x_1, x_2, \dots \, x_k$ are sufficiently close to each other, then the chief contribution to the correlator comes from coherent states situated close enough to each other, where the EFT inner product $\abs{\braket{z|z'}} = e^{- \frac{\abs{z-z'}^2}{2}}$ holds. However the path integral goes over the space of all coherent states, including the ones far away. Consequently the corrections to the correlators due to these faraway coherent states is $\text{O} \lc e^{-\mathcal{N}}\rc$, i.e. the deviation from the expected vanishing inner product. Thus the $k$ point correlator in \eqref{kpt} is modified upto small corrections by
\beq \label{kpt2}
\braket{\psi_{BH}|\, O(x_1) \,O(x_2) \dots \, O(x_k)|\psi_{BH}} =  \text{Tr} \lc \rho \, O(x_1) \,O(x_2) \dots \, O(x_k)\rc + \text{O} \lc e^{-\mathcal{N}}\rc
\eeq
On a related note, from a statistical viewpoint, Hawking's paradox is resolved since almost all states in a statistical system closely mimic thermality \cite{lloyd2013pure}. These states obey a surprising property: the $k$ point correlator over such states differ from thermal correlators by a small correction in the microcanonical ensemble:
\beq \label{kpt3}
\braket{ O(x_1) \,O(x_2) \dots \, O(x_k)} =  \text{Tr} \lc \rho_{\Omega} \, O(x_1) \,O(x_2) \dots \, O(x_k)\rc + \text{O} \lc e^{-S}\rc
\eeq
where $\rho_{\Omega}$ denotes the microcanonical density matrix and $S$ denotes the entropy of the statistical system. Our analysis of coherent states culminating in equation \eqref{kpt2} therefore provides a justification why such a statistical (more precisely, kinematical) statement works in the first place. Further, the corrections in \eqref{kpt3} are consistent with our description since for a black hole $S \propto \mathcal{N}$. We note here that these small corrections are of a different nature as compared to corrections arising due to backreaction. In contrast to backreaction, our small corrections can give an O$(1)$ contribution to physical observables under certain circumstances as shown in the next subsection.

\subsection{Implications for bags of gold paradox}
\begin{figure}

\centering
\includegraphics[width=.35\textwidth]{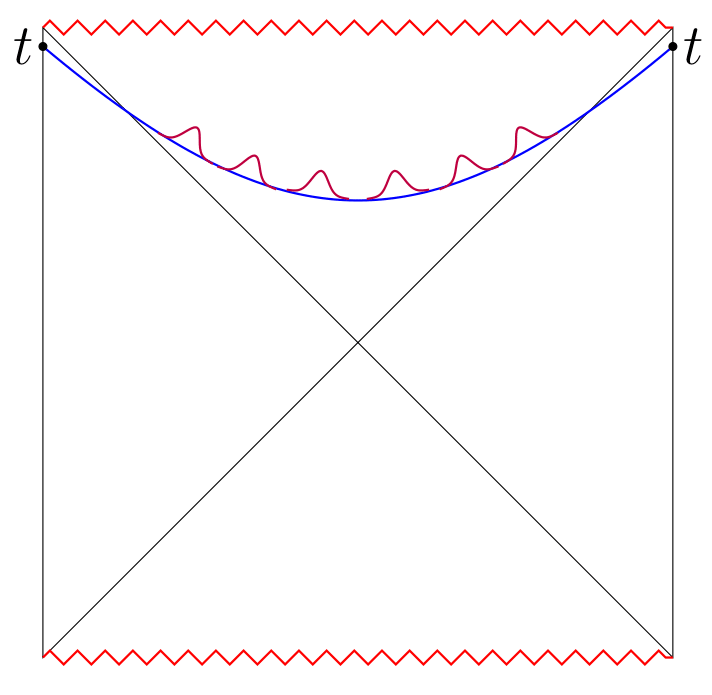}\hfill

\caption{The bags of gold paradox for eternal BH, where bumps denote excitations spatially separated far apart from each other on an interior slice whose volume increases with boundary time $t$.}
\label{paradoxfig}

\end{figure}

For correlators with operator insertions at points $x_1, x_2, \dots \, x_k$ spaced far apart from each other, there is a significant contribution to the path integral due to the non-decaying inner products between far-separated coherent states, since they represent dominant contributions. Such corrections can potentially add up to give an O$(1)$ contribution to correlators. Indeed it was shown that the spectral form factor and two-point correlator $\braket{O(t)O(0)}$ obey the usually expected decay at a small-time $t$; however, they increase and plateau off at a sufficiently large time \cite{Cotler:2016fpe}. 

Now consider a spacelike slice within the interior of a large $AdS$ black hole which has an increasingly large volume as the slice is time evolved. We place semiclassical excitations each with small support far apart from each other on these slices such that their spatial overlap is essentially zero, as shown in Figure \ref{paradoxfig}.  Within EFT, these states have a vanishing inner product when the distance between them is sufficiently large. Therefore, an estimation of the black hole's entropy using EFT leads to a substantial overcounting, far more significant than the Bekenstein-Hawking entropy. Further, such an EFT treatment of excitations leads to a violation of the spectral observables such as the spectral form factor as well as of the two-point correlator  $\braket{O(t)O(0)}$ \cite{Chakravarty:2020wdm}.  

 These inconsistencies are resolved by accounting for small inner products between far-separated coherent states in the presence of gravity \cite{Chakravarty:2020wdm, CR}. Without invoking our present discussion of path integrals, it was shown there that the entropy, spectral form factor, and other thermodynamic observables do not give wayward estimates when such small corrections are adequately taken into account. Our present discussion of path integrals is a reformulation of this picture, which provides a straightforward and intuitive understanding of how such puzzles originate starting from first principles.

\section{Conclusion}
Thus, we have provided a qualitative sketch of how small corrections can help resolve various paradoxes involving black holes from the bulk viewpoint. A significant problem is understanding why the coherent states in the presence of gravity have a fat tail of overlaps from the bulk perspective. Other important direction is to address the role of small corrections in the context of thermalization, and to understand how much of our analysis carries over to flat space. We will leave these questions to future work.

 \begin{acknowledgments}
I thank Diksha Jain, R. Loganayagam, Akhil Sivakumar and especially Suvrat Raju for various important discussions leading to this work. I acknowledge gratitude to the people of India for their steady and generous support to research in basic sciences.

\end{acknowledgments}

\bibliographystyle{JHEP}
\bibliography{refer.bib}
\end{document}